\documentclass[prd,singlecolumn,nofootinbib,showkeys,citeautoscript]{revtex4}
\pdfoutput=1
\usepackage{hyperref}
\usepackage{graphicx}
\usepackage{epstopdf}
\usepackage{amsmath}
\usepackage{dcolumn}
\usepackage{multirow}
\usepackage{fancybox}
\usepackage{bm}
\usepackage{multirow}
\usepackage{array}
\usepackage{rotating}
\usepackage{ulem,fancyvrb}
\usepackage{latexsym}
\usepackage{amsfonts}
\usepackage{xcolor}
\usepackage{longtable}
\usepackage{amssymb}
\def\tana2{\tan^2\alpha_H}
\def\tanb2{\tan^2\beta}

\def\m0pr{m_0'}

\def\Mz2{M_Z^2}

\def\met100{\slashed{E}_T\geq 100~{\rm GeV}}

\newcommand{\beqn}{\begin{eqnarray}}
\newcommand{\eeqn}{\end{eqnarray}}
\newcommand{\be}{\begin{equation}}
\newcommand{\ee}{\end{equation}}

\def \n34{\tilde{\chi}^{0}_{3,4}}

\def\125{$\sim 125$ GeV~}

\def\st{Stueckelberg~mechanism~}

\allowdisplaybreaks[1]

\begin{document}

\title{Perspectives on Higgs Boson   and Supersymmetry}

\author{Pran~Nath} 
\email{nath@neu.edu}
\affiliation{Department of Physics, Northeastern University,
 Boston, MA 02115-5005, USA}

\begin{abstract}
We review the recent discovery of the Higgs like particle at $\sim 125$ GeV  and its 
 implications for particle physics models. Specifically the implications of  the
 relatively high Higgs mass for the  discovery of supersymmetry are discussed. 
 Several related topics such as naturalness and supersymmetry, dark matter and unification
 are also discussed.
 \end{abstract}


\maketitle

\section{Introduction}

Using the combined 7 ~TeV and 8~TeV data, the ATLAS ~\cite{:2012gk} and CMS~\cite{:2012gu}
Collaborations find a signal for a Higgs like boson as follows:
ATLAS Collaboration finds a signal at a mass of 
$126.0 \pm0.4 ({\rm stat})\pm 0.4({\rm sys})~{\rm GeV}$ at the $5.0\sigma$ level while 
CMS Collaboration finds a signal at $125.3\pm 0.4 ({\rm stat})\pm 0.5({\rm sys})~{\rm GeV}$
at the $5.0\sigma$ level.  While there is the general belief that the observed particle is the 
long sought after Higgs boson  ~\cite{Englert:1964et,Higgs:1964pj,Guralnik:1964eu} 
of the electroweak theory~\cite{Weinberg:1967tq,salam}, 
 its properties still need to be experimentally established. 
 Assuming it is a Higgs boson then its mass fits very well  in the  supergravity grand unification 
 model with radiative breaking of the electroweak symmetry 
  (SUGRA)~\cite{Chamseddine:1982jx,Nath:1983aw,Hall:1983iz,Arnowitt:1992aq,Ibanez:2007pf} 
    which predicts the Higgs boson mass to be 
  less than 
 130 GeV~\cite{Akula:2011aa,Akula:2012kk,Arbey:2012dq,Ellis:2012aa}
 (for a previous review see~\cite{Nath:2012nh}).  
Also the analysis within  
SUGRA model shows that the
 high mass of the Higgs boson implies that the CP odd Higgs boson mass 
   $m_A >300$ GeV and so one is in the decoupling limit (see also Ref.~\cite{Maiani:2012qf}).
Of course as mentioned above one still needs to experimentally establish the spin and CP properties 
of the boson. 
Since the observed boson does decay into two photons, the Landau-Yang theorem~\cite{Yang:1950rg}
forbids
the particle to have spin one, but it could be spin 0 or spin 2. If it were a spin zero particle,
which is most likely the case, one still needs to establish its CP properties, i.e., the particle
could be CP even or CP odd. 
The spin and CP properties can be established by an analysis of data using the processes 
$pp\to h^0\to ZZ\to l_1 \bar l_1 l_2 \bar l_2$
and $pp\to h^0\to W^+W^- \to l^+l^-\nu\bar \nu$ (see, e.g.,~\cite{Choi:2002jk}  and \cite{Ellis:2012wg} and the references therein).
However, even after it spin and CP properties are established further work is required to 
identify the observed particle as the Higgs boson ~\cite{Englert:1964et,Higgs:1964pj,Guralnik:1964eu,Guralnik:2009jd}
which is responsible for the breaking of the
electroweak symmetry and generating masses  for the ordinary quarks and leptons via
spontaneous breaking.\\

Thus, for example, if the discovered  boson was indeed a Higgs boson of the Standard Model 
electroweak theory, then its couplings to fermions and to dibosons will have the form 
$
L_{h^0A\bar A} = -\frac{m_f}{v} h^0 f\bar f  - \frac{2M_W^2}{v}  h^0 W^{\mu +} W_{\mu}^- - \frac{M_Z^2}{v} 
 h^0 Z^{\mu}Z_{\mu} + \cdots$.  A hint of new physics will emerge from deviations of the Higgs couplings 
 from the above predictions. Thus one may define the ratio 
$R_{h^0A\bar A} = g_{h^0 A\bar A}^{\rm BSM}/g_{h^0 A \bar A}^{\rm SM} \,$, and one  can parameterize 
$R_{h^0A\bar A}$ as follows $R_{h^0A\bar A} = 1 + \Delta_{A}$, where any deviation from the standard
model are encoded in $\Delta_{A}$. Specifically one would need to experimentally determine several
couplings such as $ hb\bar b, ~ht\bar t, ~h\tau\bar \tau, ~hWW, ~hZZ, ~h \gamma\gamma, ~h  Z\gamma$
(see, e.g.,  Ref. \cite{Gunion:1989we}).
There are some hints of deviations from the Standard Model prediction in the $\gamma\gamma$ channel.
Thus  the CMS and ATLAS Collaborations give~\cite{:2012gu,:2012gk}:
$
 R_{\gamma \gamma} \equiv \hat \mu \frac{\Gamma(h\to\gamma\gamma)_{obs}}{ 
 \Gamma(h\to \gamma\gamma))_{SM}} 
= 1.8\pm  0.5 ~({\rm ATLAS }), ~1.6\pm 0.4~({\rm CMS}) \, ,
$
where
$
  \hat \mu \equiv \frac{\sigma(pp\to h)_{obs}}{\sigma(pp\to h)_{SM}} 
= 1.4\pm 0.3 {\rm ~(ATLAS)}, ~0.87 \pm 0.23 {\rm ~(CMS)} \,
$.
Now the $\gamma\gamma$ final state in the decay of the Higgs boson arises at the loop level 
 from the $W^+W^-$ in the loop and from the   $t\bar t$  in the loops. A part of the W-loop contribution,
 which is the larger of the two contributions,  is cancelled by the t-loop contribution.  Additional
 light particles are needed in the loop to produce a significant correction and several works 
 have appeared along these lines in the literature ~\cite{Carena:2011aa,Carena:2012xa,Giudice:2012pf,Joglekar:2012vc,ArkaniHamed:2012kq,An:2012vp,Abe:2012fb,Chang:2012tb}.
However, there is also the possibility that the observed excess in in fact a consequence of just the  
QCD uncertainties~\cite{Baglio:2012fu}.
It is estimated that at LHC14 with an integrated luminosity of 3000fb$^{-1}$  
one will be able to measure the couplings of the Higgs with fermions 
and with dibosons with an accuracy in the range 10-20\%~\cite{Peskin:2012we} 
while  at ILC at $\sqrt s =1$ TeV and an integrated luminosity of 1000fb$^{-1}$ 
 one should be able to measure the couplings to about 5\% accuracy~\cite{Peskin:2012we}. 
 An observation of any significant deviation will be of considerable interest in the exploration of new
 physics beyond the Standard Model.

\section{Higgs Boson and  supersymmetry}

In the Standard Model the Higgs boson mass can have a very wide range.
Thus prior to the LHC measurements,   the Higgs boson mass was constrained on the lower
side by the LEP data and on the upper side by constraints of unitarity~\cite{Dicus:1992vj,Lee:1977yc}. 
Inclusion of stability of the vacuum provides further constraints.  The analysis of vacuum stability is 
very sensitive to the next to leading order corrections and to the top mass. 
Very recently an analysis based on next-to-next -leading order (NNLO) correction 
requires that $m_h>129.4$ GeV for the vacuum to be absolutely stable 
 up to the Planck scale~\cite{Degrassi:2012ry}. This result argues that the Higgs boson mass at $\sim$ 125 GeV
 will require new physics if one wants vacuum stability up to the Planck scale (see, however, 
 \cite{Alekhin:2012py,Masina:2012tz}).
 Several suggestions have been made 
  regarding how one may stabilize the vacuum up to the Planck scale
(see, e.g., Refs~\cite{Chamseddine:2012sw,Anchordoqui:2012fq}). However, such models would 
still be subject to the large corrections to the Higgs boson mass, i.e., $m_h^2= m_0^2+ O(\Lambda^2)$
where $\Lambda$ is the cutoff which could be $O(M_G)$ where $M_G\sim 10^{16}$GeV is the GUT mass,  requiring a fine tuning in 
$m_h^2$ to 1 part in $10^{28}$. Supersymmetry avoids this large fine tuning problem and at the same time
it does not suffer from the problem of vacuum instability.\\

Now in supersymmetric theories the Higgs boson mass is predicted to be less than $M_Z$ ~\cite{Nath:1983fp} and one needs loops
corrections to lift the mass above $M_Z$. It was predicted that  with inclusion of the loop corrections the 
Higgs boson mass  for the  mSUGRA (sometimes referred to as CMSSM)
case ~\cite{Chamseddine:1982jx,Nath:1983aw,Hall:1983iz,Arnowitt:1992aq}
 the Higgs boson mass lies  below around $\sim 130$ GeV~\cite{Akula:2011aa,Arbey:2012dq,Ellis:2012aa}. Thus it is quite remarkable that the Higgs boson mass eventually  ended up
just below the upper limit predicted by mSUGRA. 
Now getting the Higgs boson as high as around $\sim 125$ GeV requires an optimization of the loop correction. 
Thus the largest correction to the Higgs boson mass arises  from the top-stop sector and is given by
\cite{Carena:2002es,Djouadi:2005gj}
$\Delta m_{h^0}^2\simeq  \frac{3m_t^4}{2\pi^2 v^2} \ln \frac{M_{\rm S}^2}{m_t^2} 
+ \frac{3 m_t^4}{2 \pi^2 v^2}  \left(\frac{X_t^2}{M_{\rm S}^2} - \frac{X_t^4}{12 M_{\rm S}^4}\right)+\cdots$.
Here $M_{\rm S}$ is the average stop mass, 
  $v=246$~GeV ($v$ is the Higgs VEV),  and $X_t$ is given  by
$ X_t\equiv A_t - \mu \cot\beta$ where $\mu$ is the Higgs mixing parameter and $\tan\beta= <H_2>/<H_1>$
where $H_2$ gives mass to the up quarks and $H_1$ gives mass to the down quarks and leptons.
  The  maximization of the loop  correction occurs   when   
   $ X_t \sim \sqrt 6 M_{\rm S}$ and 
    achieving a Higgs boson mass of size $\sim 125$ GeV requires
    $A_t/M_{\rm S}$ to be sizable.  Thus one finds that the $\sim 125$ GeV Higgs boson leads to 
   restrictions on model building~\cite{Cao:2012yn,Cohen:2012wg,Abe:2012fb,Cheng:2012pe,Dev:2012ru,Ajaib:2012eb,Kaplan:2012zzb,Liu:2012qu,Ibanez:2012zg,Hebecker:2012qp,
Aparicio:2012iw,Li:2012jf,Boehm:2012rh,Ellis:2012xd,CahillRowley:2012rv,Feng:2012jf,Perez:2012gg,Dhuria:2012bc,Wymant:2012zp,Baer:2012up,Baer:2012uya,Chakraborti:2012up}.
 It should, of course, be clear that in the above analysis we are not discussing global supersymmetry
 since breaking of supersymmetry is difficult in globally supersymmetric theories. Further, in globally
 supersymmetric theories one cannot cancel the vacuum energy and we need to work in the
 framework of local supersymmetry which requires inclusion of gravity~\cite{Nath:1975nj,Arnowitt:1975xg}. 
 Thus within the supergravity
 unified framework one can obtain a viable breaking of supersymmetry as well as arrange for the
 vacuum energy to cancel~\cite{Chamseddine:1982jx,Nath:1983aw,Hall:1983iz}. 
 Recently  mechanisms for small vacuum energy  have been discussed~\cite{Sumitomo:2012wa}
 within a class of string models such as, e.g.,~\cite{Acharya:2008zi,Conlon:2005ki}.

\subsection{Naturalness and supersymmetry}
As discussed above supersymmetry provides a natural solution to the big 
hierarchy problem, i.e, a natural cancelation between the quark loops and squark
loops that evades the need for a cancellation of   1 part in  $10^{28}$ GeV. However, 
breaking of the electroweak symmetry induced by soft parameters appears to bring in 
another smaller hierarchy 
 or the so called little hierarchy problem. Thus the electroweak symmetry breaking has the
form 
$\mu^2 = -\frac{1}{2} M_Z^2 + [{(m_{H_1}^2)- (m_{H_2}^2)\tan^2\beta}]/
{(\tan^2\beta -1)}  \,$, and if the Higgs masses and $\mu$ 
have sizes in the TeV region, then a significant cancellation is needed to 
arrange them to cancel so that $M_Z$  has the size that is seen experimentally. 
It turns out, however, that there exist regions in the parameter space of SUGRA models
where the Higgs masses can get large while $\mu$ remains small. This regime of 
radiative breaking is the Hyperbolic Branch (HB)~\cite{Chan:1997bi,Chattopadhyay:2003xi,Baer:2012uy,Baer:2012cf,Baer:2012mv}.
 Now HB has recently been classified~\cite{Akula:2011jx} 
and shown to contain Focal Points (FP)~\cite{Feng:1999mn}, Focal Curves (FC) 
and Focal Surfaces (FS).
Focal Points~\cite{Feng:1999mn}  are those regions where $m_0$ can get large while $\mu$ remains small,
Focal Curves
are those regions where $(m_0, A_0)$ (the asymptotic form of this  focal curve gives $|A_0/m_0|=1$~
\cite{Feldman:2011ud})
 or $(m_0, m_{1/2})$ can get large
while $\mu$ remains small, and Focal Surfaces are those where $m_0, A_0, m_{1/2}$
can all get large while $\mu$ remains small. It turns out that the data from the LHC 
points to most of  the parameter space consistent with all constraints lying on either
Focal Curves or Focal Surfaces. The central point of the HB branch is that one can have natural 
TeV size scalars if they lie on HB.  
In the profile likelihood analysis using the LHC data one finds that 
the scalars are typically heavy, i.e., in the TeV region, and the CP odd
Higgs mass is  greater than 300 GeV~\cite{Akula:2012kk}, which implies that one is in the decoupling limit~\cite{Haber:1994mt,Gupta:2012mi} 
which is defined so that $cos^2(\beta-\alpha)< 0.05$, where $\alpha$ is the mixing angle between
the two CP-even Higgs bosons of MSSM.
Another corroborating evidence that one is in the decoupling limit comes from the 
$B_s^0\to \mu^+\mu^-$  branching ratio
being very close to the SM value since the supersymmetric correction to 
this process proceeds by the exchange of the Higgs bosons in the direct 
channel~\cite{Choudhury:1998ze,Babu:1999hn,Bobeth:2001sq,Dedes:2001fv,Arnowitt:2002cq,Mizukoshi:2002gs,Ibrahim:2002fx}.
Thus LHCb gives 
$Br(B^0_s\to \mu^+ \mu^-) < 4.5 \times 10^{-9}$  while the standard model result is 
$Br(B^0_s\to \mu^+ \mu^-) \sim 3.2\times10^{-9}$  which shows that the supersymmetric contribution is rather
small which points to SUSY being in the decoupling limit.  We note in passing that the experimental value of 
$Br(B^0_s\to \mu^+ \mu^-)$ puts important constraints on dark matter (see Refs. \cite{Arnowitt:2002cq,Feldman:2010ke}).
However, even when $m_0$ is large one may still have many sparticles which are
light, such as the stops, the gluino, the chargino, and the neutralino~\cite{Akula:2012kk}. 
The sparticle masses are also susceptible to CP phases (for a review see~\cite{Ibrahim:2007fb}). \\

More recently an idea called natural SUSY has been discussed.
There are various versions of this idea, but in general terms it implies that the particles that enter in
the Higgs loops should be light while others could be heavy. 
Now the set of particles that participate in the Higgs boson loops are  
$(\tilde t_L, \tilde b_L), \tilde t_R, \tilde H_2, \tilde H_1$. Thus specifically one requires that the 
stops be light and they could be as light as $\sim 200$ GeV and thus could have been missed
in LHC analyses. It is certainly worthwhile to pursue possible signatures for such low mass
sparticle and strategies to this effect have been discussed in several works, see, e.g.,  
Refs.~\cite{Essig:2011qg,Allanach:2012vj}.
Finally, often EWSB fine tuning criteria are used to favor low scale SUSY. However, such criteria
may be unrealistic since there are many other sectors where fine tunings also enter such as in 
the flavor sector and a more comprehensive approach is needed to include them~\cite{Jaeckel:2012aq}.
Inclusion of these constraints tends to favor the SUSY  scale in the TeV region.\\

The anomalous magnetic moment of the muon provides an important constraint on models.
In the Standard Model the electroweak correction to $a_{\mu}=(g_{\mu}-2)/2$ 
arises from the exchange of the $W$ and of  the $Z$ boson and is estimated to be 
$\delta a_{\mu}= (28.7\pm 8.0)\times 10^{-10} $ (using $e^+e^-$ annihilation to 
estimate the hadronic error) which is $~3.6\sigma$ excess over the Standard Model prediction
and is 
$\delta a_{\mu}= (19.5\pm 8.3)\times 10^{-10} $~\cite{Hoecker:2010qn,Hagiwara:2011af}.
(using the $\tau$ decay data to estimate the hadronic error)
which is  a  $2.4\sigma$ excess over the Standard Model value. 
A more recent estimate which includes a 
 tenth-order QED contribution~\cite{Aoyama:2012wk}
and uses the hadronic error analysis 
based on $e^+e^-$ annihilation  gives   $\delta a_{\mu}= 24.9 (8.7) \times 10^{-10}$ 
which is a three  $~2.9\sigma$ excess over the Standard Model prediction. 
Now supersymmetry gives contributions to the $g_{\mu}-2$ via the exchange of the 
charginos and  sneutrinos and via the exchange of neutralinos and  smuons~\cite{Yuan:1984ww,Kosower:1983yw}. 
If the scalar masses are high, the SUSY correction tends to be small and thus 
the experiment produces a tension between theory and experiment. 
However, it is known that two loop corrections  are
significant and could reduce this tension~\cite{Heinemeyer:2004yq}.
Alternately, there could be F term or D term corrections to the Higgs boson mass~\cite{Moroi:1992zk,Babu:2008ge,Martin:2010dc,Martin:2009bg,Moroi:2011aa,Endo:2011mc,Endo:2011gy}
 thus reducing the need for a large SUSY loop correction which alleviates the tension.

\section{Supersymmetry and asymmetric dark matter}

An interesting idea concerns cosmic coincidence which relates to  the fact that ~\cite{Komatsu:2010fb}
$\frac{\Omega_{DM}h_0^2 }{ \Omega_B h_0^2} = 4.99 \pm 0.20\,$, where 
$\Omega_B h_0^2$ is the relic density of baryonic  matter and  
 $\Omega_{DM} h_0^2$ is the  dark matter relic density. The fact that the ratio is $O(1)$  implies that perhaps 
 there is a common origin to these two components (for a review see ~\cite{Davoudiasl:2012uw}).
 There is a suggestion that dark matter originates from the transfer of $B-L$ from the visible sector to 
 dark sector in the early universe  (For a recent work see ~\cite{Kaplan:2009ag}). 
 There are two main elements that such an idea involves. The first is that one needs a mechanism for 
 the transfer of $B-L$ from the visible sector to the  dark sector. The second is that the dark matter that is
 produced thermally is dissipated and does not contribute to the total dark matter density. Various 
 mechanisms have been discussed for the transfer and are mentioned in ~\cite{Feng:2012jn}.
 One example is the interaction $ L_{transfer} = M^{-3} \psi^3LH$ where $\psi$ is the dark matter
 particle which carries a lepton number and thus a non-vanishing $B-L$.  Using the constraints of 
 thermal equilibrium in the early universe~\cite{Harvey:1990qw,kolbturner}   
  one can compute the ratio $n_X/n_B$ where $n_X$ is the number of dark matter particles and 
  $n_B$ is the number of baryons which allows one to compute the ratio of the relic densities 
  of dark matter to baryons, i.e., the ratio  $\Omega_{DM}/\Omega_{B}$. \\
  
  The second issue concerns
  how one may dissipate the dark matter that is thermally produced. In order to accomplish this we
  use the Stueckelberg mechanism~\cite{Kors:2004dx,Kors:2004ri,Kors:2005uz,Feldman:2006wb,Feldman:2007wj}
  where by the dark particles that are thermally produced
  can annihilate  into the Standard Model particles via a direct channel $Z'$ pole. The $Z'$ 
    corresponds to a gauged $U(1)_X$ symmetry,
  where for $U(1)_X$ we choose $U(1)_{B-L}$.  In this case annihilation of the thermally  produced
  dark matter via  the Breit-Wigner $Z'$ boson can completely dissipate the thermally produced dark
  matter. For the supersymmetric case one has in this picture two dark matter particles:
  the neutralino and the $B-L$ carrying X particle.  In order for the cosmic coincidence to work one 
  must have most of the dark matter, i.e., about 90\%, constituted of X, while the remainder can be
  neutralino. It is then interesting to ask if the neutrino which would have only about say 10\% of the
  total relic density of dark matter could be observed in dark matter experiments. It turns out that
  indeed this is the case and the neutralino can still be detected in the current direct detection experiments 
     ~\cite{Aprile:2012nq,Aprile:2011hi,Aprile:2011hx,Aprile:2010um}  
  and  in  future experiments such as XENON-1T~\cite{Aprile:2012zx} and SuperCDMS-1T~\cite{Cabrera:2005zz}. 
 We note in passing that an interesting recent idea is the  so called dynamical dark matter 
 which is constituted of an ensemble of different particle species consisting of different  masses
 and abundances  rather than of one type. Interesting implications of such ideas have been 
 discussed in ~\cite{Dienes:2011ja,Dienes:2012yz}.

\section{Supersymmetry and unification}
 Grand unified  models typically have two problems: 
 the first concerns the fact  that in GUT models such as SO(10)  many Higgs boson fields are involved in the 
breaking of the GUT symmetry and one has to adjust the VEVs so that the breakings occurs close to each other 
and one has the SM gauge group unifying at one scale. This is specifically true of the SO(10) model on which we focus
here.  Thus, in SO(10) 
$16+\overline {16}$ or $126+\overline{126}$  is used for rank reduction, and one uses 
  $45$, $54$ or  $210$ for breaking the symmetry down to the standard model gauge symmetry.
    Further, for   the electroweak symmetry breaking one uses $10$ plets of Higgs fields. It turns out that one can 
  replace  the above array of Higgs fields by $144+\overline{144}$ which can break the SO(10)
  gauge symmetry down to $SU(3)_C\times U(1)_{em}$~\cite{Babu:2006rp,Babu:2005gx}.  
  The textures in this class of models  are different than in conventional SO(10) models
  (see, e.g., Ref. ~\cite{Chattopadhyay:2001mj}).  
  One interesting feature here ~\cite{Babu:2006rp,Nath:2009nf}  is that  a large $\tan\beta$~\cite{Ananthanarayan:1991xp}
  is not needed for $b-t-\tau$ unification in this class of models.  
  Special techniques are needed for the computation of 
  SO(10) couplings which have been developed in Refs.~\cite{Nath:2001uw,Nath:2001yj,Nath:2005bx}.  
Secondly, as is well known, grand unified models have a doublet-triplet problem, i.e., how to make all the
color Higgs triplets heavy and one pair of Higgs doublets light. One method used is the so called missing partner
mechanism which has been implemented in $SU(5)$ in  ~\cite{Grinstein:1982um,Masiero:1982fe}.
 The extension of the missing partner mechanism to  SO(10)  was done more recently and four different
 cases have  been identified \cite{Babu:2006nf,Babu:2011tw}. The phenomenology of these models needs to
 be further worked out. \\

We discuss now briefly the current status of proton stability. In supersymmetric theories proton decay can occur
via dimension 4, dimension 5 and dimension 6 operators (for reviews see Refs. \cite{Nath:2006ut,Raby:2008pd,Hewett:2012ns}). 
 Dimension 4 operators  must be forbidden because they
give  too rapid  a proton decay and this can be accomplished by R parity conservation. Dimension six operators arise
from the exchange of vector lepto-quarks and the dominant decay mode here is $p\to e^+ \pi^0$.
Proton decay modes can allow one to discriminate between GUTs and  strings~\cite{Arnowitt:1993pd}.
Specifically the mode $p\to e^+\pi^0$ can allows one, in principle,   to  distinguish between GUT models and
 D brane models ~\cite{Klebanov:2003my,Cvetic:2006iz}. 
 The current limit from Superkamiokande for this mode is ~\cite{Abe:2011ts}
 $\tau(p\to e^+\pi^0) > 1.4\times 10^{34} ~{\rm yrs}$ and it is expected that in the future at
 Hyper-K one will be able to achieve a  sensitivity~\cite{Abe:2011ts} of
   $\tau(p\to e^+\pi^0) > 1\times 10^{35}~{\rm yrs}$. This brings one to the edge of 
   observability since unified models predict a decay lifetime which is typically $10^{36\pm 1}$ yrs.\\

Proton decay from dimension five operators is the  most model dependent. 
Typically GUT models give too rapid  a proton decay and one needs either mass
suppression due to the SUSY model being located on the Hypebolic Branch of radiative
breaking of the electroweak symmetry\cite{Chan:1997bi}
 or a cancellation mechanism ~\cite{Nath:2007eg} 
  where $B\&L$ violating dimension five operators
from various sources tend to cancel.  Proton decay is also sensitive to CP phases~\cite{Ibrahim:2007fb}.
The current experimental situation is as follows: 
Super-K gives the limit
 $\tau(p\to \bar \nu K^+) > 4\times 10^{33} {\rm yrs}$, while Hyper-K in the future will reach a  sensitivity of 
$\tau(p\to \bar \nu K^+) > 2\times 10^{34} ~{\rm yrs}$. It is important to keep in mind that dimension 5 proton 
decay is very sensitive to the sparticle spectrum and  thus observation of sparticles and the measurement of 
their masses would make the proton lifetime from dimension five operators much more predictive. 
 
\section{Stueckelberg extensions\label{st}}

The \st allows generation of mass for a  vector boson which is a $U(1)$ gauge field without the
 necessity of a Higgs mechanism.  The \st arises quite naturally in compactification of extra
 dimensions and in string models where it is directly connected with the Green-Schwarz anomaly
 cancellation term.
 While the \st  had been around for a long time, its application 
 to particle physics was  made only recently~\cite{Kors:2005uz,Kors:2004iz}
 where $U(1)_X$ extensions  ($U(1)_X$ is a hidden sector gauge symmetry) 
  of the standard model and of 
 MSSM were given.  Since then numerous applications of these have been made~\cite{Kors:2005uz,Kors:2004iz,Kors:2004dx,
 Feldman:2006wd, Cheung:2007ut,Feldman:2007nf,Feldman:2006wb} 
  (for related works see, e.g., ~\cite{Cassel:2009pu,Mambrini:2009ad}). 
 These applications  arise largely because of the mixing between $U(1)_X$ and $U(1)_Y$ of the
 standard model which allows  the hidden sector to communicate with the visible sector. 
 The D\O\   Collaboration~\cite{Abazov:2010ti}  and the CMS detector Collaboration ~\cite{Chatrchyan:2012it}  
 have put new limits  on this class  of models.
 The \st has found many applications such as in the explanation of the 
 PAMELA anomaly, in the dijet anomaly and in protecting~\cite{Feldman:2011ms} R parity against violations
 due to RG running~\cite{Barger:2008wn}.

\section{Conclusion}
The observation by ATLAS and CMS of a boson of mass around 125 GeV, has important implications
for physics beyond the standard model especially supersymmetry. Although the properties of the 
new boson still need to be fully established, it is the general perception that the observed particle
is indeed the long sought after Higgs boson that enters in the electroweak symmetry breaking.
While in the Standard Model this boson could have a mass over a rather wide range, in supersymetry
it has an upper limit of about 150 GeV. Specifically, in mSUGRA the mass of the Higgs boson 
is predicted to lie below 130 GeV. It is  rather interesting that the observed boson turns out to have a mass
which does lie below this limit. An interesting issue for further exploration concerns 
the nature of the sparticle landscape~\cite{Feldman:2007zn,Feldman:2008hs,Nath:2010zj} 
under the 125 GeV Higgs boson constraint.
Further, it is now imperative that one explore the implications of this
constraint on the allowed parameter space of SUSY, string and D brane models and identify the 
lightest sparticles that lie within reach of the upgraded LHC. In summary, the observation of 
the boson, assuming it is the Higgs boson, provides strong support for supersymmetry
as it is within the framework of supersymmetry that that an elementary spin zero Higgs boson
has a natural setting. Thus the focus of the LHC now must turn to the discovery of 
supersymmetry in the next round of experiments.   

\section*{Acknowledgments}
This  work is  supported in
part by the U.S. National Science Foundation (NSF) grants 
PHY-0757959, PHY-070467, TG-PHY110015  and DOE  NERSC grant DE-AC02-05CH11231.

\end{document}